\documentclass[fleqn,usenatbib]{mnras}

\usepackage{newtxtext,newtxmath}
\usepackage{orcidlink}
\usepackage{xcolor}
\usepackage{amsmath,amstext}
\usepackage[T1]{fontenc}
\usepackage{natbib}
\usepackage{caption}
\usepackage{subcaption}
\usepackage{microtype}
\usepackage{longtable}
\usepackage{verbatim}
\usepackage{hyperref}
\hypersetup{
    colorlinks=true,
    linkcolor=blue,
    filecolor=magenta,      
    urlcolor=cyan,
}

\usepackage[T1]{fontenc}

\DeclareRobustCommand{\VAN}[3]{#2}
\let\VANthebibliography\thebibliography
\def\thebibliography{\DeclareRobustCommand{\VAN}[3]{##3}\VANthebibliography}

\usepackage{graphicx}	
\usepackage{amsmath}	

\newcommand{\hst}{\textit{HST}}

\newcommand{\jwst}{\textit{JWST}}

\newcommand{\gsim}{\gtrsim}

\newcommand{\ha}{\hbox{H$\alpha$}}

\newcommand{\grizli}{\textit{Grizli}}

\newcommand{\sfrr}{\hbox{SFRr$_{H\alpha/BB}$}}
\newcommand{\leta}{\hbox{$\log(\eta_{2300})$}}

\definecolor{aggiemaroon}{HTML}{500000}

\newcommand{\editone}[1]{\textcolor{black}{#1}}

\title[Star Formation in a Galaxy Pair at Cosmic Noon with CANUCS]{When, Where, and How Star Formation Happens in a Galaxy Pair at Cosmic Noon Using CANUCS JWST/NIRISS Grism Spectroscopy}

\author[V. Estrada-Carpenter et al.]{
Vicente Estrada-Carpenter$^{1}$\orcidlink{0000-0001-8489-2349}\thanks{E-mail: astro.vince.ec@gmail.com (VEC)},
Marcin Sawicki$^{1}$\thanks{Canada Research Chair}\orcidlink{0000-0002-7712-7857},
Gabe Brammer$^{2,3}$\orcidlink{0000-0003-2680-005X},
Guillaume Desprez$^{1}$\orcidlink{0000-0001-8325-1742},
\newauthor
Roberto Abraham$^{4,5}$\orcidlink{0000-0002-4542-921X},
Yoshihisa Asada$^{1,6}$\orcidlink{0000-0003-3983-5438},
Maru\v{s}a Brada\v{c}$^{7,8}$\orcidlink{0000-0001-5984-0395},	
Kartheik G. Iyer$^{9}$\orcidlink{0000-0001-9298-3523},
\newauthor
Nicholas S. Martis$^{1,10}$\orcidlink{0000-0003-3243-9969},
Jasleen Matharu$^{2,3}$\orcidlink{0000-0002-7547-3385},
Lamiya Mowla$^{11}$\orcidlink{0000-0002-8530-9765},
Adam Muzzin$^{12}$,
Ga\"el Noirot$^{1}$,
\newauthor
Ghassan T. E. Sarrouh$^{12}$\orcidlink{0000-0001-8830-2166},
Victoria Strait$^{2,3}$\orcidlink{0000-0002-6338-7295},
Chris J. Willott$^{10}$\orcidlink{0000-0002-4201-7367}
\\
$^{1}$Institute for Computational Astrophysics and Department of Astronomy \& Physics, Saint Mary's University, 923 Robie Street, Halifax, NS B3H 3C3, Canada\\
$^{2}$Cosmic Dawn Center (DAWN), Denmark\\
$^{3}$Niels Bohr Institute, University of Copenhagen, Jagtvej 128, DK-2200 Copenhagen N, Denmark\\
$^{4}$David A. Dunlap Department of Astronomy and Astrophysics, University of Toronto, 50 St. George Street, Toronto, Ontario, M5S 3H4, Canada\\
$^{5}$Dunlap Institute for Astronomy and Astrophysics, 50 St. George Street, Toronto, Ontario, M5S 3H4, Canada\\
$^{6}$Department of Astronomy, Kyoto University, Sakyo-ku, Kyoto 606-8502, Japan\\
$^{7}$Department of Mathematics and Physics, Jadranska ulica 19, SI-1000 Ljubljana, Slovenia\\
$^{8}$Department of Physics and Astronomy, University of California Davis, 1 Shields Avenue, Davis, CA 95616, USA\\
$^{9}$Columbia Astrophysics Laboratory, Columbia University, 550 West 120th Street, New York, NY 10027, USA\\
$^{10}$National Research Council of Canada, Herzberg Astronomy \& Astrophysics Research Centre, 5071 West Saanich Road, Victoria, BC, V9E 2E7, Canada\\
$^{11}$Whitin Observatory, Department of Physics and Astronomy, Wellesley College, 106 Central Street, Wellesley, MA 02481, USA\\
$^{12}$Department of Physics and Astronomy, York University, 4700 Keele St. Toronto, Ontario, M3J 1P3, Canada\\
}

\date{Accepted XXX. Received YYY; in original form ZZZ}

\pubyear{2023}

\begin{document}
\label{firstpage}
\pagerange{\pageref{firstpage}--\pageref{lastpage}}
\maketitle

\begin{abstract}
Spatially resolved studies are key to understanding when, where, and how stars form within galaxies.  Using slitless grism spectra and broadband imaging from the CAnadian NIRISS Unbiased Cluster Survey (CANUCS) we study the spatially resolved properties of a strongly lensed ($\mu$ = 5.4$\pm$1.8)  z = 0.8718 galaxy pair consisting of a blue face-on galaxy (10.2 $\pm$ 0.2 log($M/M_\odot$)) with multiple star-forming clumps and a dusty red edge-on galaxy (9.9 $\pm$ 0.3 log($M/M_\odot$)).  We produce accurate H$\alpha$ maps from JWST/NIRISS grism data using a new methodology that accurately models spatially varying continuum and emission line strengths.   With spatially resolved indicators, we probe star formation on timescales of  $\sim$ 10 Myr (NIRISS H$\alpha$ emission line maps) and $\sim$ 100 Myr (UV imaging and broadband SED fits). Taking the ratio of the H$\alpha$ to UV flux ($\eta$), we measure spatially resolved star formation burstiness.  \editone{We find that in the face-on galaxy both H$\alpha$ and broadband star formation rates (SFRs) drop at large galactocentric radii by a factor of $\sim$ 4.7 and 3.8 respectively, while SFR over the last $\sim$ 100 Myrs has increased by a factor of 1.6. Additionally, of the 20 clumps identified in the galaxy pair we find that 7 are experiencing bursty star formation, while 10 clumps are quenching and 3 are in equilibrium (either being in a state of steady star formation or post-burst). Our analysis reveals that the blue face-on galaxy disk is predominantly in a quenching or equilibrium phase. However, the most intense quenching within the galaxy is seen in the quenching clumps.}  This pilot study demonstrates what JWST/NIRISS data can reveal about spatially varying star formation in galaxies at Cosmic Noon.
\end{abstract}

\begin{keywords}
galaxies: Evolution -- galaxies: Formation
\end{keywords}



\begin{figure*}
 \includegraphics[width=\linewidth]{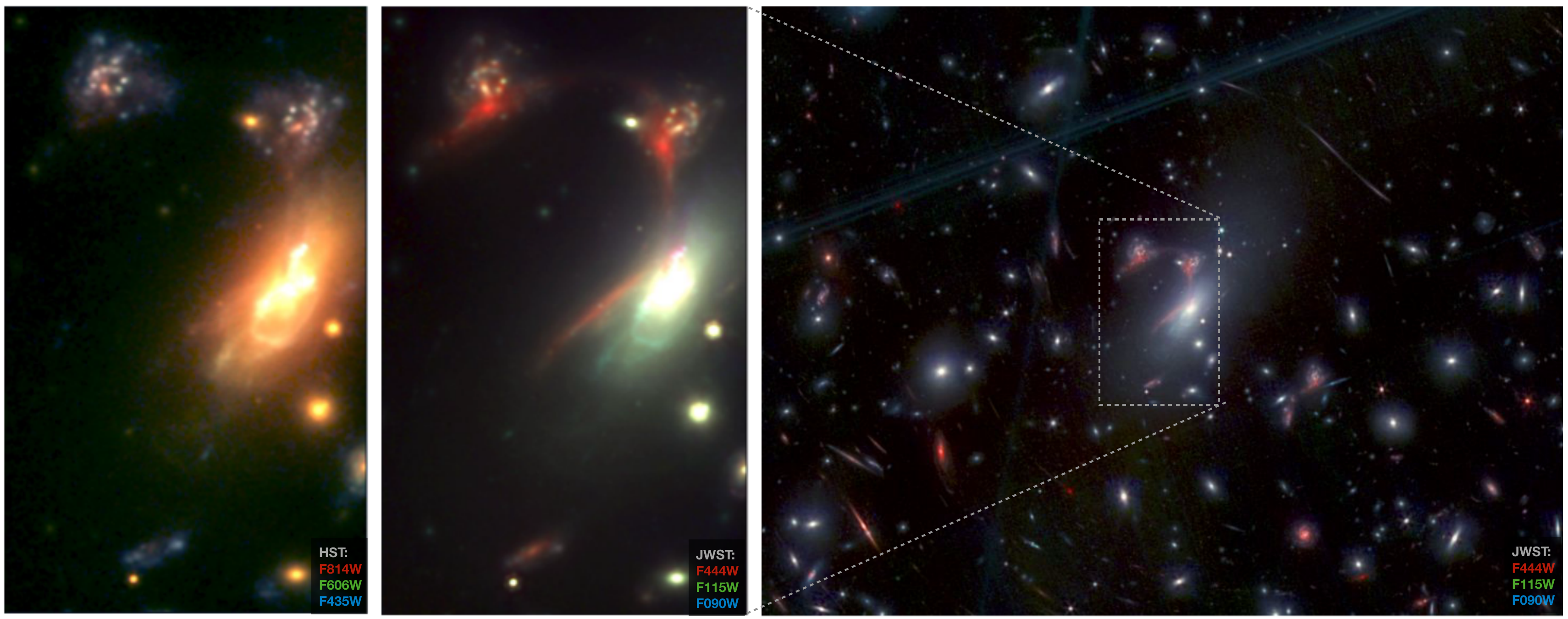}
 \caption{False colour images of the Question Mark Pair and MACS J0417.5-1154 (right panel). The left two panels are zoom-in images of four of the multiply-lensed images of the Question Mark Pair taken with HST and JWST. By comparing the JWST and HST images we see how dusty the red edge-on galaxy is as it is barely visible in the HST/ACS imaging.}
 \label{fig:eye}
\end{figure*}

\section{Introduction}
Knowing when, where, and how star formation occurs within galaxies is crucial to understanding how galaxies grow, evolve, and quench. A powerful way to constrain when star formation occurs is to look at star formation rate (SFR) indicators over various timescales. Observable indicators such as H$\alpha$ and rest-frame UV flux have been used to constrain SFRs \citep{kenn94, kenn12} at lookback times of $\sim$ 100 Myrs and $\sim$ 10 Myrs respectively. The difference in the SFR timescales is due to where the flux originates from: H$\alpha$ emission line flux is caused by short-lived O and B-type stars, while rest-frame UV flux is dominated by longer-lived stars. By comparing shorter and longer SFR timescales we can piece together the recent star-formation history within galaxies. 

Several spatially-unresolved studies have used ratios of H$\alpha$ and rest-frame UV flux to estimate how bursty a galaxy's star formation is (\cite{glaz99, igle04, lee09,lee11,weis12,domi15, meht17, brou19, emam19, feth21, brou22, meht23, asad23}, Asada et al. in prep). The ratio of H$\alpha$ to rest-frame UV flux (hereafter referred to as $\eta$) works as a burstiness indicator as H$\alpha$ flux responds more quickly to changes in the star-formation history (SFH) than does rest-frame UV flux. Using SFHs to calibrate $\eta$ we can determine whether a galaxy is bursting with star formation, is in an equilibrium state, or its SFR is decreasing. 

One way to study where star formation occurs in galaxies is to exploit spatially resolved photometric and spectroscopic data. Studies using integral field units (IFUs) have researched the spatially resolved star formation properties of galaxies at z < 0.2 \citep{fern13,gonz16,belf18,medl18,salv19,avil23}. These works have studied topics such as inside-out quenching, spatially resolved star formation, the star-forming main sequence of galaxies, and specific star-formation rate (sSFR) profiles. \editone{IFUs have also offered spatially resolved insights at redshifts ranging from 1 to 2, utilizing KMOS \citep{stot14,beif17} and SINFONI \citep{schr11,moli17}. These studies have explored various topics, including star formation in clumpy star-forming regions and the impact of feedback, the formation ages of massive galaxies, the correlation between metallicity gradients and sSFR, and \ha\ SFRs.} Our goal is to do a similar analysis at Cosmic Noon ($z\sim 1-3$) using JWST/NIRISS slitless grism data. Among the key strengths of the James Webb Space Telescope (JWST, \citealt{gard23}) are its exceptional sensitivity and spatial resolution of the spectroscopic modes. These attributes enable the spectroscopic observations of distant galaxies with an unprecedented level of detail, surpassing what was previously possible with other facilities, including the Hubble Space Telescope (HST). Imaging studies conducted with JWST have already revealed the remarkable intricacies present in the structure of distant galaxies (e.g., \citealt{mowl22, abdu23,gime23,asad23}). However, the capabilities of JWST extend beyond broadband or medium-band imaging alone: JWST is equipped with the Near-Infrared Imager and Slitless Spectrograph (NIRISS, \citealt{doyo12, doyo23}) whose Wide Field Slitless Spectroscopy (WFSS, \cite{will22}) mode captures spatially-resolved low-resolution \editone{(R $\sim$ 150)} spectroscopic information from all resolved objects in a 2.2\arcmin$\times$2.2\arcmin field all at once.

Previous works have used HST \citep{nels16, math20, math22, meht23} and JWST/NIRISS slitless grism spectroscopy \citep{math23} to study the spatial distribution of star formation. These works used a method of extracting emission line maps that ignores the spatially varying nature of galaxy spectral energy distributions. Specifically, until now, it has been common to assume that the underlying stellar populations do not vary across the extent of the galaxy -- an assumption that is understandable in the case of low-SNR data. However, JWST/NIRISS observations call for the development of more advanced data analysis techniques to fully harness the immense potential offered by this new instrument. Particularly pressing is the need for a more comprehensive treatment when investigating target galaxies that show spatial variations in their spectral energy distributions (SEDs). The need for an improved approach was recognized before the launch of JWST through simulations that revealed potential biases in emission line maps and integrated line flux measurements \citep{sorb17}. Our goal has thus been to develop such an improved approach to extracting information from spatially-resolved slitless grism spectroscopy. 

In this paper, we present a demonstration of our proof of concept analysis of the spatially resolved star formation properties of the Question Mark Pair (QMP), a complex system of two z=0.8718 galaxies that we observed with NIRISS slitless spectroscopy and NIRCam imaging behind the lensing cluster MACS J0417.5-1154 as part of the CAnadian NIRISS Unbiased Cluster Survey (CANUCS). Figure~\ref{fig:eye} highlights our target. The rightmost panel of Figure \ref{fig:eye} shows a false colour image of MACS J0417.5-1154 using JWST/NIRCam filters, with zoomed-in images of the QMP shown using HST/ACS filters (leftmost) and JWST/NIRCam (centre). These images show four (of a total of five) lensed images of the QMP and highlight the complex structure of the pair of galaxies, with one of the galaxies being a heavily dust-obscured, edge-on galaxy and the other a face-on disk-like system with strings of star-forming clumps. In Section 2 we give more information about the target galaxy pair and describe the data we used from the CANUCS survey. In Section 3 we discuss spatially resolved modelling of the NIRISS grism data. In Section 4 we discuss our results including the derived emission line map. In Section 5 we analyze the spatially resolved properties of the QMP.  In Section 6 we summarize our conclusions. Throughout this work, we use AB magnitudes and assume a cosmology with $\Omega_{m,0}$ = 0.3, $\Omega_{\lambda,0}$ = 0.7, and $H_{0}$ = 70 km s$^{-1}$.

\section{Data and Target}
\begin{figure*}
 \includegraphics[width=\linewidth]{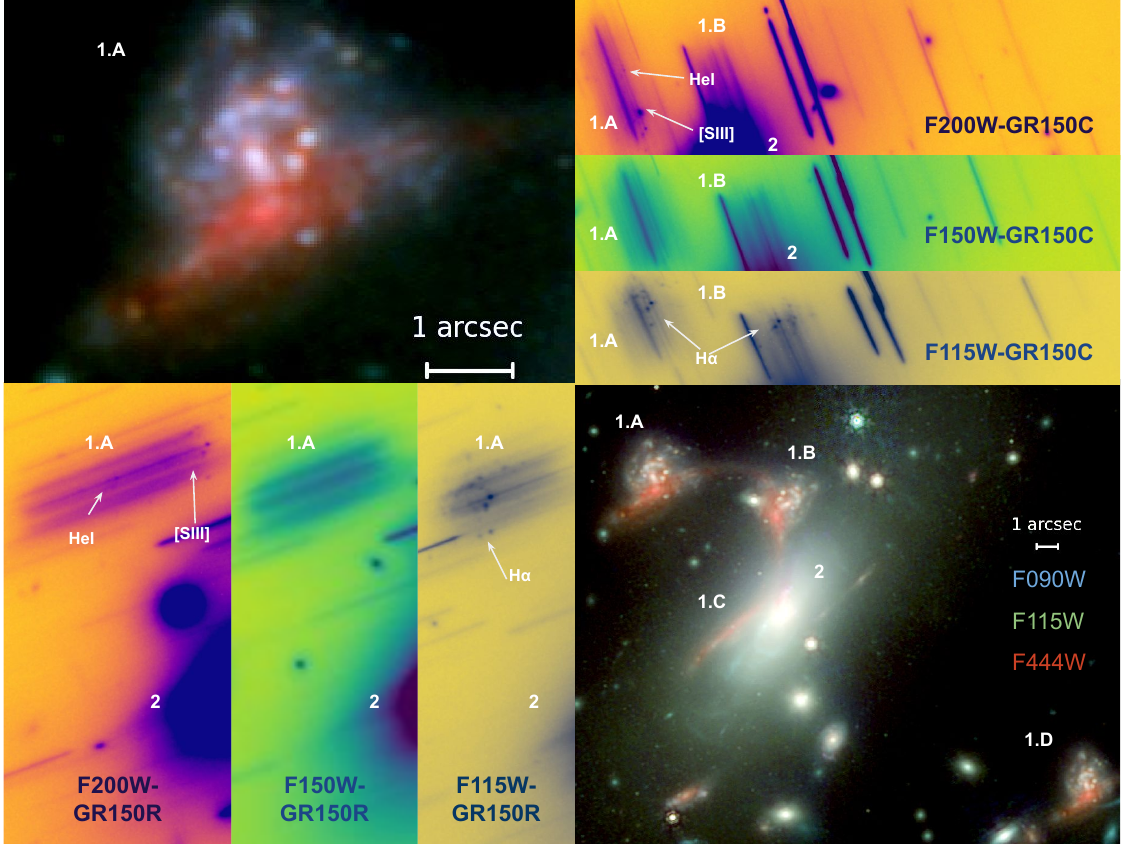}
 \caption{JWST/NIRISS grism data and false colour images. In each of the panels, we have marked objects of interest. The bottom right panel shows the multiply imaged QMP (1.A-D) in the MACS J0417.5-1154 cluster, here we see four of the five images of the galaxy, the fifth being under the BCG (2). The top left panel shows the image of the QMP we will be examining as it is the most magnified and least contaminated of the multiple images. The top right and bottom left panels show the NIRISS grism data in all orients and filters, focusing on the image of the QMP we are using.}
 \label{fig:data}
\end{figure*}
\subsection{Target: The Question Mark Galaxy Pair}

Our objective is to study spatially resolved star formation properties derived from JWST/NIRISS emission line maps. The QMP is an excellent case study as this complex system has several clumpy star-forming regions, is composed of two potentially interacting galaxies with different stellar populations, and has increased SNR due to it being highly lensed. 

To analyze spatially resolved properties from JWST slitless grism data we developed a new methodology for the recovery of emission line maps.  To demonstrate the value and effectiveness of our new method, we decided to demonstrate it on the QMP, which is a particularly complex galaxy system. 

Figure~\ref{fig:data} shows the  NIRCam images of the target of our study along with its NIRISS grism data. In each of the panels we have marked several objects of interest, including objects 1.A-D which are 4 of the 5 lensed images of the QMP, the 5th is obscured by the $z=0.441$ brightest cluster galaxy (BCG) which is labelled as object 2. The top right and bottom left panels show the NIRISS grism data (in all three filters) of the field near the QMP. Here we can clearly see H$\alpha$ emission in the F115W data for the QMP (1.A for GR150R and 1.A and 1.B in GR150C marked by an arrow). The bottom right panel shows more of the field along with the reason we refer to this galaxy system as the QMP as the red galaxy (not previously seen in the HST data) outlines a question mark, though it should be noted that the point of the question mark is an unrelated galaxy. The QMP system consists of two galaxies that are magnified and slightly distorted by the gravitational potential of the $z=0.441$ foreground galaxy cluster MACS J0417.5-1154: a blue face-on galaxy that consists of a disk and several clumpy star-forming regions and a highly dust-obscured red edge-on galaxy. The face-on galaxy in our pair has been previously reported by \cite{jauz19} to be a multiply-lensed ``complex ring galaxy'' at $z=0.8718$ and called ``The Doughnut'' by these authors. Our deep JWST/NIRCam imaging reveals that this galaxy is in fact not a classic ring galaxy but a disk sprinkled with star-forming clumps. This face-on disk galaxy is accompanied by a previously unknown edge-on red galaxy that is at a spectroscopically confirmed similar redshift (Sec. \ref{sec_ha_map}). We dubbed the galaxy pair ``The Question Mark Pair'' given their prominence in what looks like a question mark-shaped multiply-imaged lens system traced by the red galaxy in our NIRCam imaging (see bottom-right panel of \editone{Figure \ref{fig:data}}).

We focus our attention on the top-left image of the QMP (Figure \ref{fig:data}) top left panel), which corresponds to Image 1.1 in \citealt{jauz19}. We do so because this image is the most magnified among the multiple images of the QMP ($\mu$=5.4$\pm$1.8 in the \citealt{jauz19} lens model), while at the same time not suffering from strong shear, as can be seen by comparing its appearance with those of the other images of the galaxy pair in Figure \ref{fig:data}, and the least contaminated in the NIRISS grism data. By focusing on this complex, highly-resolved system we highlight the main benefits of high-SNR spatially resolved NIRISS grism spectroscopy and showcase our new line-map extraction method. 

\begin{figure*}
 \includegraphics[width=\linewidth]{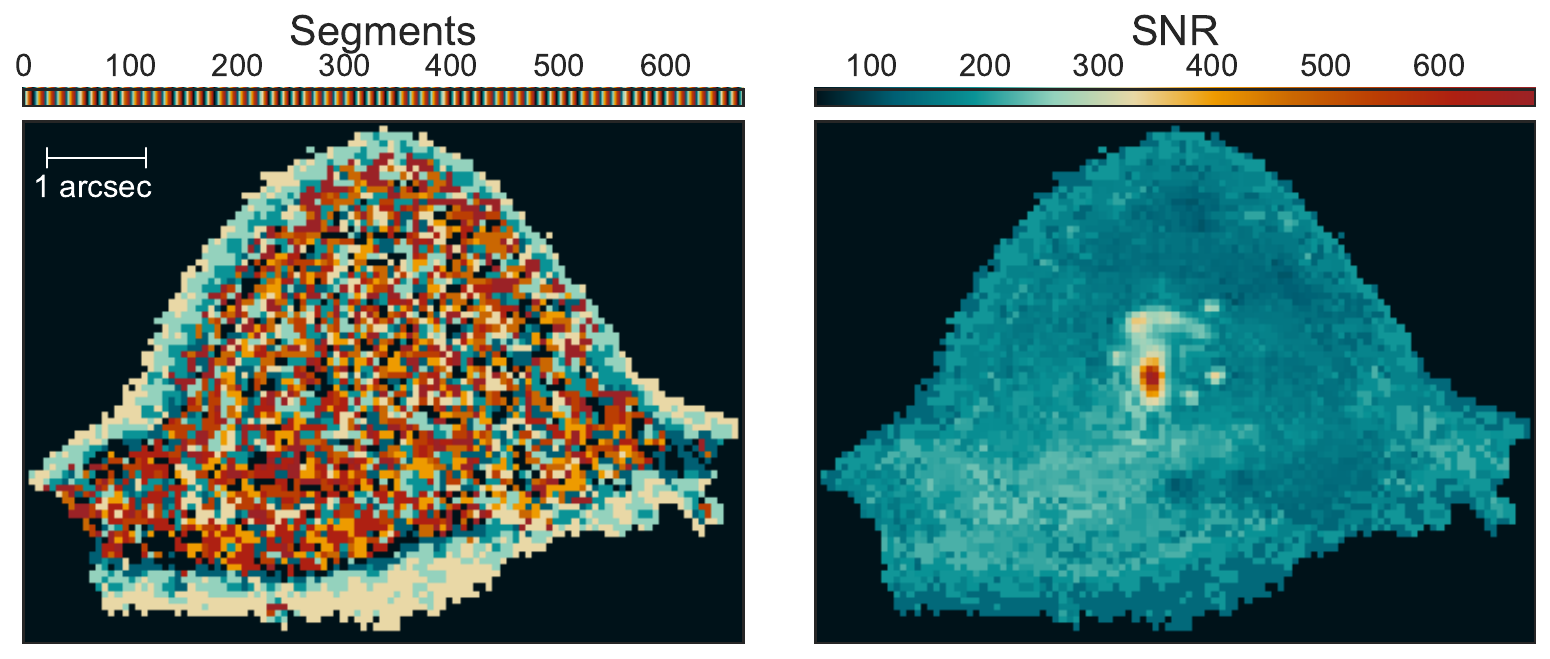}
 \caption{Segmentation map for the QMP. The first panel shows the 675 segments that the system has been broken into. In the centre of the system most of the segments are only 1 pixel and towards the outer regions (where the SNR is the lowest) we have larger regions. On the right, we show the SNR map of the system where the SNR has been calculated for each segment of our segmentation map. Here we see that in the inner region, many of the single-pixel regions sit above our SNR limit while the outer regions sit at the SNR=100 limit.}
 \label{fig:seg}
\end{figure*}

\subsection{Data}

We use data from the MACS J0417.5-1154 cluster field taken as part of the CAnadian NIRISS Unbiased Cluster Survey (CANUCS, \cite{will22}). From this data set, we utilize broadband imaging in \jwst\ NIRCam (F090W, F115W, F150W, F200W, F277W, F356W, F410M, F444W) with exposure times of 6.4 ks each with SNR between 5 to 10 for an AB = 29 point source \citep{asad23,stra23}. In addition, we use archival \hst\/ACS data (F435W, F606W, F814W). MACS J0417.5-1154 cluster galaxies were modelled and removed from the imaging (Martis et al. submitted). \editone{The Point Spread Functions (PSFs) for broadband imaging were empirically determined by median-stacking non-saturated bright stars. As a result, our broadband data consist of PSF-convolved images standardized to the resolution of the JWST NIRCam F444W.}

The \jwst\ NIRISS grism data were taken in both the GR150R and GR150C grism utilizing the F115W, F150W, and F200W filters with exposure times of 19.2 ks in all filters \citep{math23}. The grism data were first processed using the grism modelling analysis software \grizli\ \citep{grizli} which performs an end-to-end reduction of the data and modelling of sources. Subsequent processing of the data was performed after the 2nd, 1st, 0th, and -1st orders of the cluster galaxy spectra were modelled and removed (see Sec \ref{sec:clst_remv})

\section{Methods}
\subsection{Segmentation \label{sec_seg}}

When using broadband imaging, spatially resolved studies can treat each pixel or group of pixels as an independent object and fit galaxies pixel by pixel or region by region \citep[e.g.,][]{abra99, wuyt12, sawi12, sorb15, sorb18, abdu23}. However, due to the self-contaminating nature of grism data \citep{vand11, estr19} the elements cannot be treated as independent along the angle of dispersion. Therefore a pixel-by-pixel fit of the galaxy with the grism data would require that all elements be fit simultaneously. There are thus two main limitations to a pixel-by-pixel fit with grism data. The first of these is computing resource limitations. When fitting, each element must be forward-modelled in each exposure for each filter, and so, depending on the size of the galaxy and the number of exposures, we can easily exhaust our computing resources. The second main limitation is signal-to-noise (SNR). Pixels in the outskirts of the galaxy have low SNR and fitting them will likely only return the priors. Therefore it would be best to group lower SNR pixels to create larger/higher SNR regions that can be fit to derive more robust fits while simultaneously reducing the computation time and resources used \citep{wuyt12, wuyt13, wuyt14, tada14, lang14, chan16, feth20}. 

For our spatially resolved analysis we segment the galaxies into regions, grouping pixels using colors and fluxes from the broadband images. To do this, we first take the brightest pixel in the direct NIRISS F150W image of the galaxy and using a nearest neighbours algorithm find the distance to all pixels in color space using all possible combinations of our 11 broadband filters. We then start to group the nearest neighbours of our brightest pixel until we reach an SNR limit (based on the NIRISS F150W SNR map); note that for the brightest pixels, this will often be satisfied with the single pixel alone. Here we do not consider positional information and regions do not have to be contiguous. For this proof of concept study, we decided to fit with as many regions as possible therefore we chose an SNR sufficient to create 675 regions (SNR = 100), which was the number that we could fit with our available computing resources.  Once we have reached the SNR limit for our region we mask those pixels from the broadband images and start the process over again using the newest brightest pixel. We repeat this process until all pixels have joined a region. Note that the last region added will always sit below our SNR limit, but this will likely not affect our fits as this region will consist of low SNR background pixels. \editone{In the process of developing our nearest neighbours method, we conducted tests against other segmentation methods that utilize Gaussian mixture models and Voronoi tessellation. Our findings indicated that our nearest neighbours method generated regions with more homogeneous properties, such as lower scatter in sSFR, dust, \ha\ equivalent width, and so on. Furthermore, contamination modelling performed with our nearest neighbours segmentation resulted in more accurate models.}

\begin{figure*}
 \includegraphics[width=\linewidth]{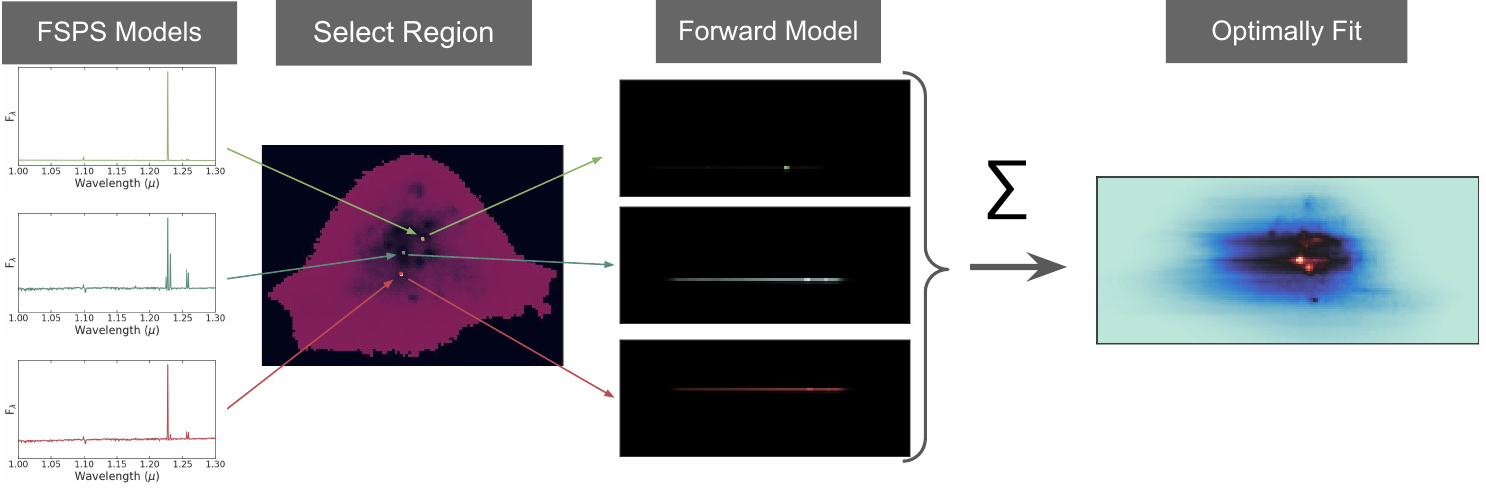}
 \caption{Schematic of the forward-modelling procedure for the multi-region analysis method. First, we show a subset of the input spectra from FSPS we used to model, here we utilize 4 spectra per region which results in a total of 4$\times$675 = 2700 spectra. We then forward-model each region using our segmentation map for each filter in both orients, resulting in 675$\times$3$\times$2 = 129,600 forward-modelled grism spectra. We then sum the linear combination of the models using coefficients that optimize the fit resulting in the complete forward-modeled grism spectrum. For simplicity, only a single orient in one filter (F115W) is shown here, although our procedure simultaneously models all three filters in both orients. }
 \label{fig:fwd_mdl}
\end{figure*}

Figure \ref{fig:seg} shows our segmentation of the QMP. In the left panel of Figure \ref{fig:seg} we see the results of our segmentation. In the centre of the galaxy, we have many single-pixel regions (274, $\sim$ 40$\%$ of our regions) with an overall median region size of 2 pixels and our largest region being 494 pixels. The right panel of Figure \ref{fig:seg} shows the SNR map of our regions. Here we see that most of the galaxy is shaded blue, which sits at our SNR limit of 100, with several regions sitting well above this limit (most of these regions being our 1 or 2-pixel regions). 

\subsection{Spatially Resolved Priors / Broadband Fits \label{sec_BB}}
Segmenting the galaxy into several regions results in a large parameter space we need to explore. \editone{Therefore to streamline the fitting process we derive a set of bespoke priors for each region by fitting to the spatially resolved broadband images.}

We start by using our PSF-matched NIRCam and ACS images, then apply the segmentation we derive in Section \ref{sec_seg}.  The Flexible Stellar Population Synthesis code (FSPS) \citep{conr10} is used with a combination of MILeS and BaSeL libraries and a Kroupa initial mass function \citep{krou01}.

We fit each of the 675 regions using a set of 50,000 pre-generated FSPS models \editone{(which include nebular lines)} varying sSFR, t$_{50}$, both metallicities, ionization, and Av, with redshift set to z$_{spec}$ = 0.8718 \citep{mahl19,jauz19} using the nonparametric SFHs described in \cite{iyer17, iyer19}. From these fits we derive posteriors for each parameter for each region. \editone{These provide maps of physical properties (such as stellar mass etc.) and will also serve as priors for our fits to the grism data.}

\subsection{Spatially Resolved Grism modelling
\label{sec:SpatiallyResolvedmodelling}}

The goal of our spatially resolved methodology is to generate highly accurate models of the grism spectra. As our end goal is to generate the purest emission line maps possible we must model the continuum and each of the emission lines accurately. To do this, each region will be fit with a linear combination of models, allowing for complex spectra to be fit relatively quickly.

Figure \ref{fig:fwd_mdl} is a schematic of how our forward-modelling works:
\begin{itemize}
    \item Step 1) Sample from the priors of each region and create multiple models using FSPS. The number of models generated for each region depends on multiple factors but ultimately is limited by computing memory. For the QMP, we generate 4 models per region. In the leftmost panel of Figure \ref{fig:fwd_mdl} we show one of the models we may use for just three regions.
    \item Step 2) We forward-model each model spectrum \citep{estr23} for each region in all orients using grizli. For the QMP that will result in 2700 model grism spectra. The middle panels of Figure \ref{fig:fwd_mdl} show three regions being forward-modelled in one filter and one orientation using the spectra from Step 1. 
    \item Step 3) We find the linear combination of the model grism spectra which optimally fit the grism data, as seen in the rightmost panel of Figure \ref{fig:fwd_mdl} (for a single filter).
    \item Step 4) We then explore our parameter space by repeating Steps 1--3 thousands of times to sample the priors well, once our library of models is built we find the best-fit model that minimizes $\chi^2$.
\end{itemize}

\subsection{Cluster Galaxy Removal \label{sec:clst_remv}}
\begin{figure*}
 \includegraphics[width=\linewidth]{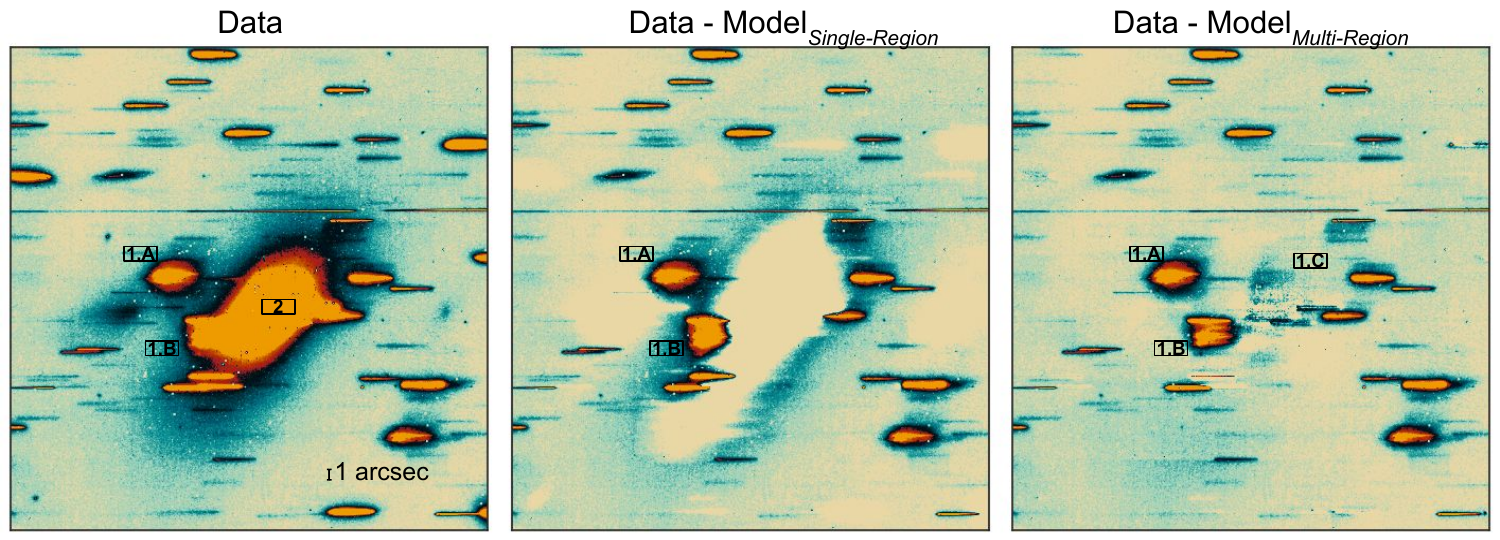}
 \caption{Cluster galaxy removal from the grism data. In each of the panels objects of interest are labeled, these are the same labels as seen in Figure \ref{fig:data}. In the left panel we show the raw data in NIRISS F115W GR150C, the middle plot shows the NIRISS data with the cluster galaxies removed using the standard single-region method of modelling contamination with polynomials, and the right panel shows the NIRISS data with the cluster galaxies removed using our spatially resolved fitting. Here we see that our spatially resolved models do a much better job of removing the contaminating sources while leaving behind sources that were previously contaminated by the cluster galaxies (object 1.C).
 }
 \label{fig:bcg_mdl}
\end{figure*}
One of the difficulties in working with grism data is dealing with contamination, i.e. spectra from other objects that overlap the spectrum we are interested in (this may come from any order of dispersion). In the MACS J0417.5-1154 data, this is made even more difficult with the presence of large/extended foreground cluster galaxies. We do gain magnification by observing in the cluster field but the trade-off is heavily contaminated grism data. The left panel in Figure \ref{fig:bcg_mdl} shows a cutout (26 arcsec $\times$ 26 arcsec) of the MACS J0417.5-1154 grism data centred on the $z=0.441$ brightest cluster galaxy (BCG). As the cluster galaxies are so bright, we have to not only worry about the 0th and 1st-order dispersions but also the 2nd and -1st orders as well. 

In the centre panel of Figure \ref{fig:bcg_mdl} we show the same field but with the single-region models for the cluster galaxies removed. Here, following the standard approach, single-region cluster galaxy models are generated by forward-modelling the galaxy spectra using a set of polynomials (up to and including 3rd order polynomials) using the image of the galaxy within the segmentation. The result is that the standard single-region models tend to over-subtract the contaminating spectra and fail to model the extended light, as seen particularly in the field centre. This over-subtraction is likely due to spatially varying stellar populations that are not accounted for in a single-region approach.  Given the limitations of the standard procedure, a better approach is needed.

To properly subtract the cluster galaxies in the grism data we use our multi-region modelling approach of Sec.~\ref{sec:SpatiallyResolvedmodelling}. To model the cluster galaxies we use isophotal models of the BCGs produced by Martis et al. in prep. These isophotal models replace our direct images and we use them to segment the galaxy and produce the forward-modeled grism spectra. We fit these bright foreground galaxies using our multi-region modelling procedure (Sec~\ref{sec:SpatiallyResolvedmodelling}) and subtract them from the NIRISS grism data.  The results of this can be seen in the right panel of Figure \ref{fig:bcg_mdl}. Our models show a clear visual improvement, including reducing over-subtraction and recovery of underlying spectra. In Figure \ref{fig:bcg_mdl} we have labelled objects of interest using the same numbering scheme from Figure \ref{fig:data}. We see in the centre of the right panel that we can recover grism spectra of galaxies that were completely contaminated by the BCG. Object 1.C is a highly deformed image of the QMP which sits close to the BCG, as seen in Figure \ref{fig:data}. Using the standard contamination subtraction method Object 1.C's spectrum is unrecoverable, but using our multi-region methods not only do we recover the spectrum but we can also produce emission line maps from the data. Our cluster galaxy multi-region grism models provide a much better contamination removal than the standard process. From our modelling, we see that the BCG added $\sim$ 12$\%$ contaminating flux to the QMP grism spectra which we successfully removed.

\subsection{Emission Line Maps}

Our goal is to produce as accurate an emission line map as possible to study the spatially resolved star formation properties of the QMP. To achieve this we must model the continuum well in addition to any nearby emission lines present within the galaxy spectra. Doing so is necessary because the way we produce an emission line map from the grism data is to first forward-model the galaxy using its best-fit model, excluding the line we are interested in extracting. We then subtract the best-fit model (excluding the line we are extracting) from the grism data and dither the resulting residuals (which will include the two orthogonal orients) to produce the emission line map. 

Our multi-region approach allows for a spatially varying continuum and emission line strength, something that is not done using the standard method of forward-modelling the entire galaxy as a single entity. Our approach is superior since not accounting for a spatially varying continuum/emission line strength may lead to over or under-subtraction of the continuum/nearby emission lines leading to missing/excessive flux in the emission line maps, as we discuss next.

\begin{figure*}
 \includegraphics[width=\linewidth]{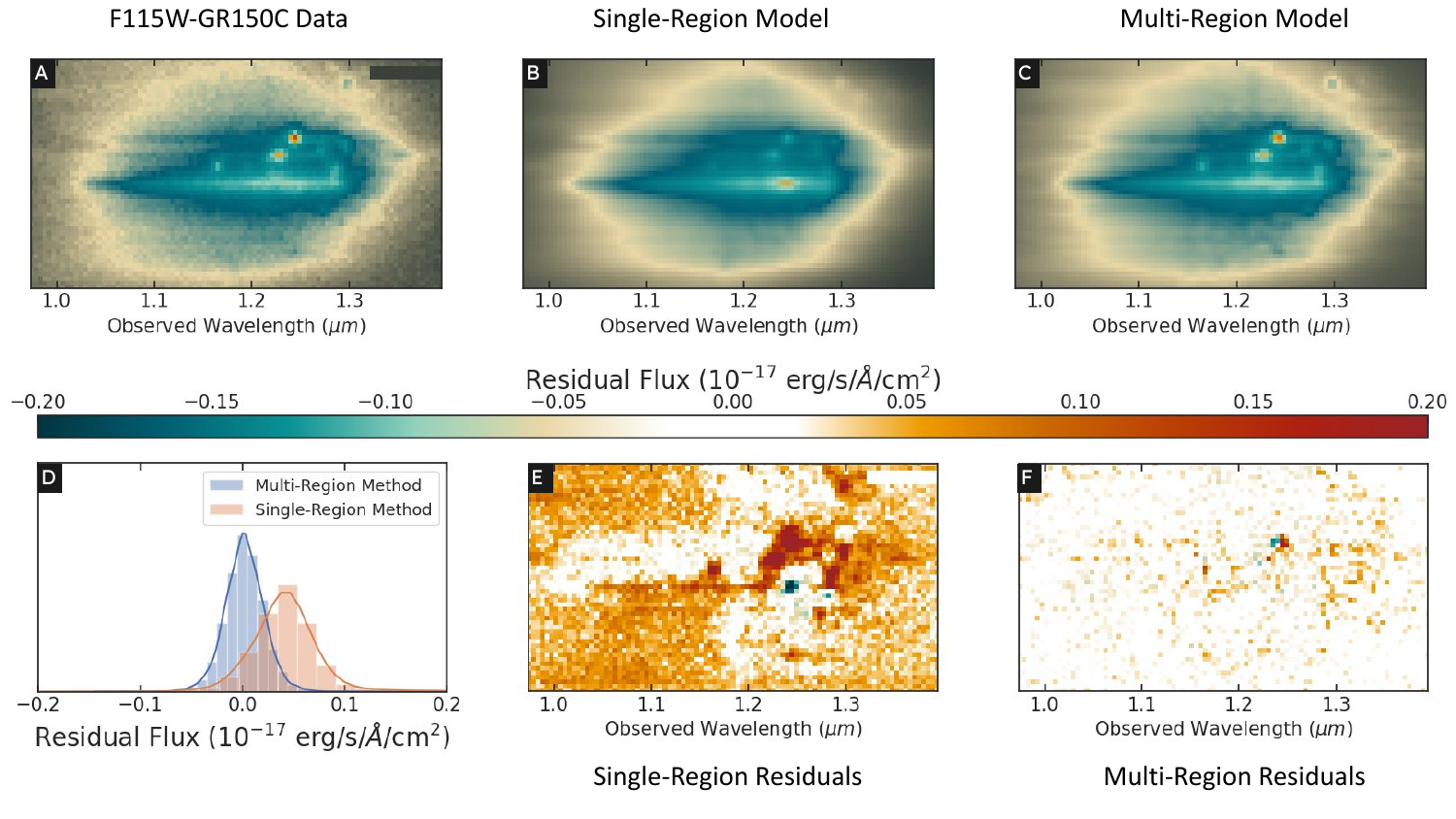}
 \caption{Panel A shows JWST/NIRISS F115W-GR150R grism data for the QMP. Panel B shows the best-fit model using the standard single-region approach, and Panel C shows our best-fit model using our multi-region approach. In the top panels, we see that our multi-region model more accurately reproduces the NIRISS grism data.  Panel D shows the multi-region residual distributions for the single-region method (orange) and multi-region method (blue); here we see that the multi-region method residuals sit at 0.0$^{+0.02}_{-0.02}$ (10$^{-17}$ erg/s/\AA/cm$^2$) - 3.5$^{+9.5}_{-1.0}$ $\%$ error - while the single region method produces offset residuals that sit at 0.04$^{+0.03}_{-0.03}$ (10$^{-17}$ erg/s/\AA/cm$^2$) - 12.9$^{+30.7}_{-3.9}$ $\%$ error -. Panel E shows the single-region residuals and Panel F shows the multi-region residuals. Here we see structure in the single-region residuals stemming from emission lines and continuum, while we see almost no structure in the multi-region residuals.
}
 \label{fig:mdl_comp}
\end{figure*}

Figure \ref{fig:mdl_comp} shows an example of our multi-region fitting versus the standard single-region fitting using the best-fit models for both methods. For the standard single-region method, grizli uses a set of optimally fit templates (13 FSPS models and 32 emission lines) to model the galaxy with the redshift set to z = 0.8718. Panel A shows the F115W-GR150C grism data for the QMP, Panel B shows the best-fit model from the standard single-region method, and Panel C shows the best-fit model from our multi-region method. From these panels, we see that the multi-region models do an excellent job of modelling the real grism spectrum both in the continuum and emission lines, while the single-region model does not do as well at modelling the emission lines and the continuum. This is the most apparent in the H$\alpha$ emission as since the single region method weighs the forward-modelled spectrum by the entire image of the galaxy, the best-fit model for the single region method contains a large amount of H$\alpha$ flux in the bulge of the blue face-on galaxy. This does not match what we see in Panel A. On the other hand, the multi-region best-fit model shows a more similar distribution of H$\alpha$ flux to the actual data (Panel A). Panels E and F show the residual flux for the single-region and multi-region approaches respectively. In Panel F we see no discernible structure within the residuals while in Panel E we see a considerable amount of residual in both the emission lines (both H$\alpha$ + NII and SII) and along several regions where the continuum has been under-subtracted (both of which will negatively affect the emission line maps). Panel D shows the distribution of the residual flux with the multi-region residuals shown in blue and the single-region residuals shown in orange. We see in panel D that the multi-region residuals show a smaller range than those of the single-region model;  they are also clustered symmetrically around zero, whereas those for the single-region model are systematically offset in the positive direction. In other words, the multi-region residual distribution is consistent with simple shot noise, whereas residuals from the single-region model show a systematic bias. Numerically, the multi-region residuals show a residual flux of 0.0$^{+0.02}_{-0.02}$ (10$^{-17}$ erg/s/\AA/cm$^2$), or a fractional error of  3.5$^{+9.5}_{-1.0}$ $\%$.  In contrast, the single-region residuals show a residual flux of 0.04$^{+0.03}_{-0.03}$ (10$^{-17}$ erg/s/\AA/cm$^2$), or a fractional error of 12.9$^{+30.7}_{-3.9}$ $\%$. It is clear that the multi-region method does better. This overall improvement means our multi-region models will contain much less contamination and will be truer to the actual spatial distribution of H$\alpha$

\begin{figure*}
 \includegraphics[width=\linewidth]{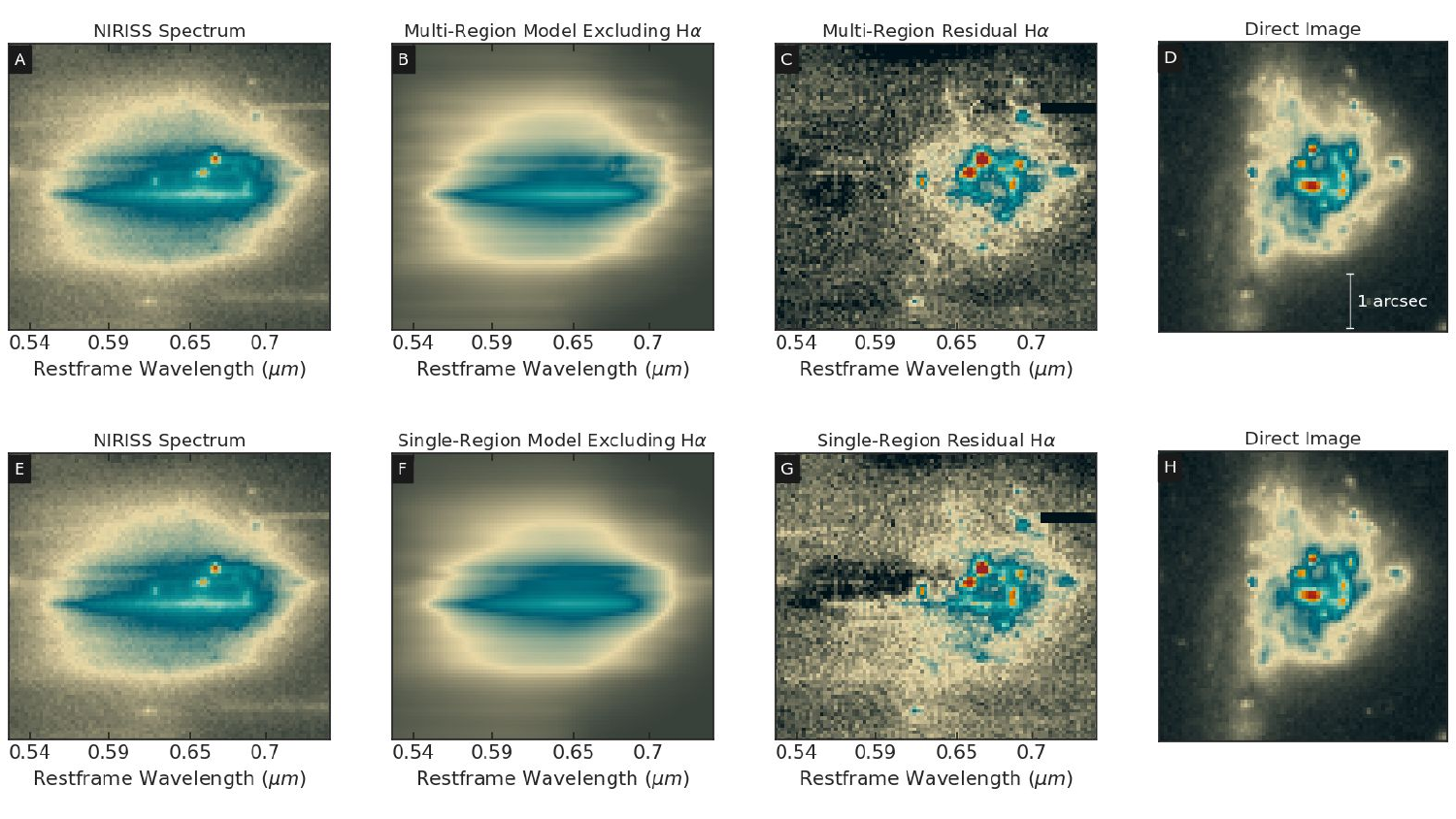}
 \caption{H$\alpha$ residuals for the multi (top row) and single (bottom row) region models. In each row the first panel is the NIRISS F115W-GR150C data for the QMP, the second is the best-fit model excluding H$\alpha$ emission, the third panel shows the residual flux when subtracting panel 2 from panel 1 in each row, and the last panel shows the NIRISS F150W direct image of the galaxy in the orient of the data for reference. Subtracting panel 2 from panel 1 should result in only H$\alpha$ flux being left behind. When comparing the multi-region (Panel C) to the single-region (Panel G) H$\alpha$ residuals we see that the multi-region residuals are very reminiscent of the QMP while the single-region residuals show extended H$\alpha$ flux which, as it does not match what we see in the final panels, we can only conclude is due to under-subtraction of the continuum flux. 
}
 \label{fig:Ha_res}
\end{figure*}

The difference in the two models leads to large differences in the emission line maps. Figure \ref{fig:Ha_res} shows the first few steps to creating an emission line map, and shows the large difference in the two models. The first panel in each row of Figure \ref{fig:Ha_res} shows the NIRISS F115W-GR150C data for the QMP. The second panel in each row shows the best-fit model for the multi-region (Panel B) and single-region (Panel F) approach excluding H$\alpha$ emission. By subtracting the second panel from the first we should be left with only H$\alpha$ emission flux, this is what is shown in the third panel on each row. Ideally, the residual flux should appear as a stamp of the QMP in H$\alpha$ emission, the final panel in each row shows the direct image of the QMP in the orient of the grism data and this is what the residual flux should resemble. In Panel C we see that the residual H$\alpha$ emission using the multi-region approach does match our expectations as the flux is distributed like the QMP. However, in Panel G we see that for the single-region approach, we are left with an excess of continuum flux that was not properly subtracted. This will result in an extended H$\alpha$ emission profile that is not really there. When we complete the process and derive the final emission line maps, we find that the standard single-region method produces a map with 50$\%$ more flux than our multi-region method. This extra flux is likely not from the target H$\alpha$ emission line, but instead is an artefact of the under-subtracted continuum and poorly modelled SII emission, all of which we see in Figure \ref{fig:mdl_comp} Panel E and Figure \ref{fig:Ha_res} Panel G. This excess flux will end up in the emission line maps which results in a 40$\%$ higher SFR in the blue face-on galaxy and 70$\%$ higher SFR in the red edge-on galaxy than our multi-region results. The differences in the SFR and flux may distort any result stemming from the standard single-region method.

\section{Physical Property Maps}

In this section, we employ our multi-region line-extraction method to investigate star formation in the QMP, following the successful demonstration of its superiority over the conventional one-region approach.

\subsection{The Question Mark Pair \label{sec:qmp}}

Figure \ref{fig:maps} shows several property maps derived either from stellar population fits to the broadband fluxes region by region (see Sec \ref{sec_BB}) or the extracted H$\alpha$ emission line map (Panel D). Panel A has been corrected for lensing while Panels D, F, G, H, and I have been corrected for lensing and dust. Note that our maps are not in the source plane, but values such as distance, mass, flux, and SFR have all been corrected for using our lens model described in the Appendix. We see in Figure \ref{fig:maps} Panel E that the two galaxies have drastically different colours with the red edge-on being highly reddened and the blue face-on galaxy having multiple colours likely indicating varying stellar populations. 

What makes the QMP so interesting is that these galaxies are possibly at the beginning of an interaction (as their morphologies do not seem to be disturbed). An interaction between the galaxy pair could lead to a burst of star formation, and this may be the reason why the blue face-on galaxy contains so many clumpy star-forming regions. To break down what the different components of the two galaxies are doing, we segment the galaxies into the regions shown in Figure \ref{fig:maps} Panel B, where we have split the galaxy pair into the blue galaxy bulge and disk, red bulge and disk, and bursting, equilibrium, and quenching clumps. \editone{We identify clumpy regions in the QMP by first applying a Gaussian smoothing algorithm to the NIRCam F115W image to create a smooth model of the galaxy. We then subtract this model from the original image leaving behind clumpy regions, note that this will also include bulges and patchy regions of the galaxy. We then employ Source Extractor for Python \citep{bert96, barb16} to segment the residual image. The mass map (see Figure \ref{fig:maps} Panel A) is utilized to identify the two galaxy bulges. Then, to eliminate patchy/noisy regions, we impose an \ha\ SNR limit of 10, this should leave only clumpy star-forming regions in our segmentation. The remaining clumpy regions are then categorized into bursty, equilibrium, and quenching phases using the method detailed in Section \ref{sec:burst}. Finally, we distinguish between the blue disk and red disk using a colour-based classification utilizing a Gaussian mixture model.}

We know that the two galaxies are at the same redshift as they both contain clumpy star-forming regions with H$\alpha$ emission at the same observed wavelength. In addition, these two galaxies appear associated, as in the three unobscured lensed images of the QMP (Figure \ref{fig:data} objects 1.A, 1.B, and 1.D) the two galaxies have the same alignment; if the two galaxies were just a chance projection from different redshifts then the two would align differently in the 3 lensed images. 

Using the stellar mass surface density map and segmentation map (Figure \ref{fig:maps} Panels A and B respectively) we have determined that this would be a major merger with a stellar mass ratio of $\sim$ 2:1 with the higher mass blue face-one galaxy having a stellar mass of 10.2 $\pm$ 0.2  log(M/M$_\odot$) and the lower mass red edge-on galaxy having 9.9 $\pm$ 0.3 log(M/M$_\odot$). 

\subsection{\ha\ Map Corrections \label{sec:corr}}
\editone{One of the main objectives of this work is to quantify the recent (10 Myr) SFR from the \ha\ flux. However, the flux in our \ha\ emission line maps is not purely from star-forming HII regions and therefore we cannot simply convert it to SFR. In order to extract the SFRs from our \ha\ map we need to apply several corrections: we correct for dust attenuation, contamination of the \ha\ line by the neighbouring [NII] line, and for the contribution of non-star-forming sources of \ha\ flux.}

\begin{figure*}
 \includegraphics[width=18cm]{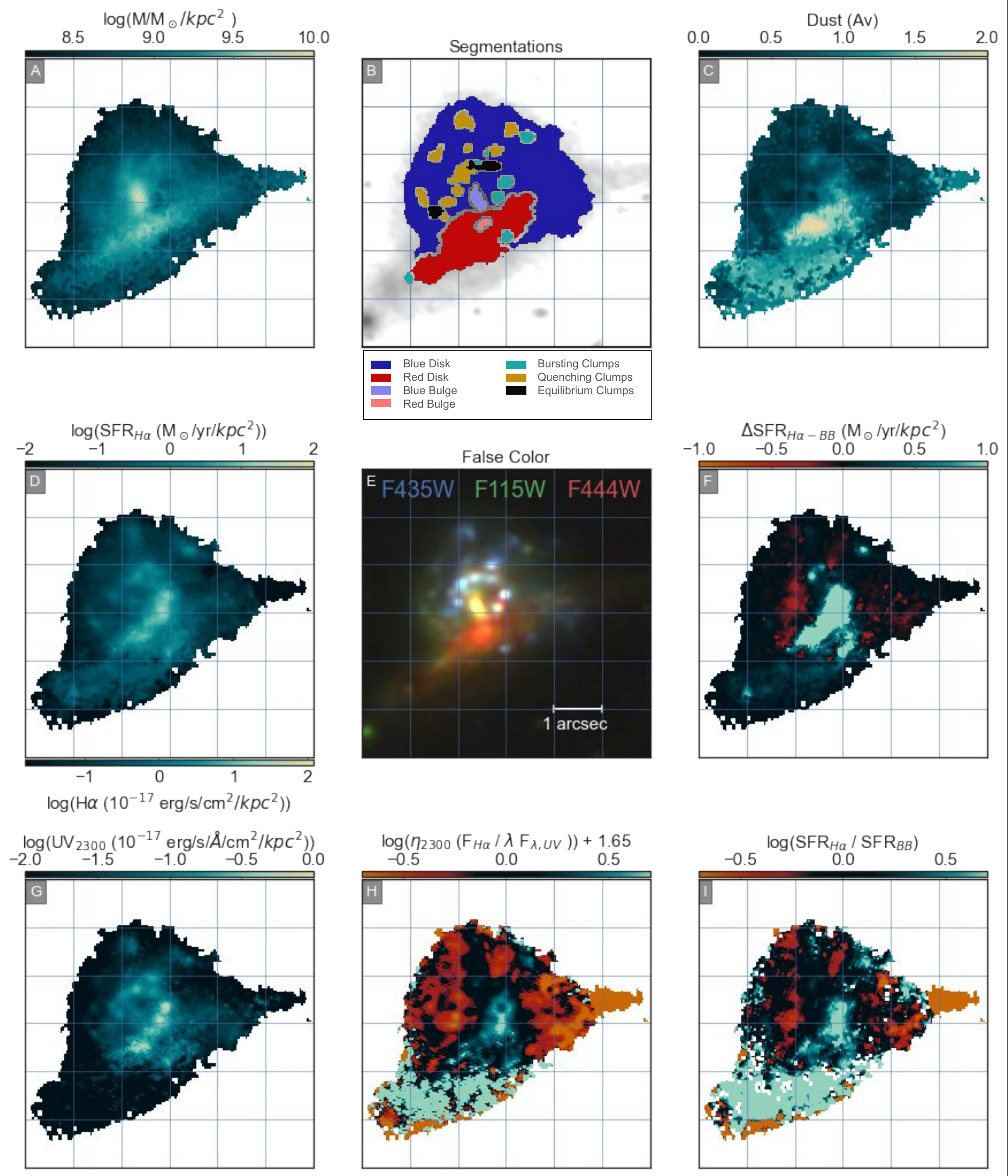}

 \caption{Maps of the stellar mass surface density, segmentation of regions plus the H$\alpha$ emitting clump in the red galaxy, dust (Av), H$\alpha$ emission line flux surface density and SFR$_{H\alpha}$ surface density, False colour image, Change in SFR from the broadband SFRs to the H$\alpha$ SFRs, UV$_{2300}$ surface density, \leta\ + equilibrium offset (1.65), and the SFR$_{H\alpha}$ to SFR$_{BB}$ ratio. Panel A has been corrected for lensing, while Panels D, F, G, H, and I have been corrected for lensing and dust.}
 \label{fig:maps}
\end{figure*}

\editone{Many of the property maps we are going to analyze can be heavily affected by dust, so before we begin our analysis we correct for dust attenuation using dust measurements from the spatially resolved broadband data. In Figure \ref{fig:maps} Panel C we show the attenuation map derived from spatially resolved SED fits to the broadband images (see Section \ref{sec_BB}). In Panel C we see a region of high attenuation ($>$ 1 A$_V$), which coincides with the red edge-on galaxy ($1.2^{+0.4}_{-0.7}$ A$_V$). In contrast, the blue face-on galaxy is less attenuated ($0.6^{+0.5}_{-0.2}$ A$_V$). We note that as our A$_V$ measurements are from the broadband data they do not account for the possibility that the extinction towards the nebular regions may be higher than the continuum \citep{char00}. To better account for the differences in dust attenuation we use the relationship E$_{stars}$(B-V) = (0.44 $\pm$ 0.03) E$_{gas}$(B-V) \cite{calz97} for our dust corrections to \ha. With the attenuation map for the QMP in hand, we apply a position-dependent dust correction to our images of interest. We do this using the Calzetti dust law \citep{calz00}, as this is the dust parameterization used in our SED fitting.} 

\begin{figure*}
 \includegraphics[width=18cm]{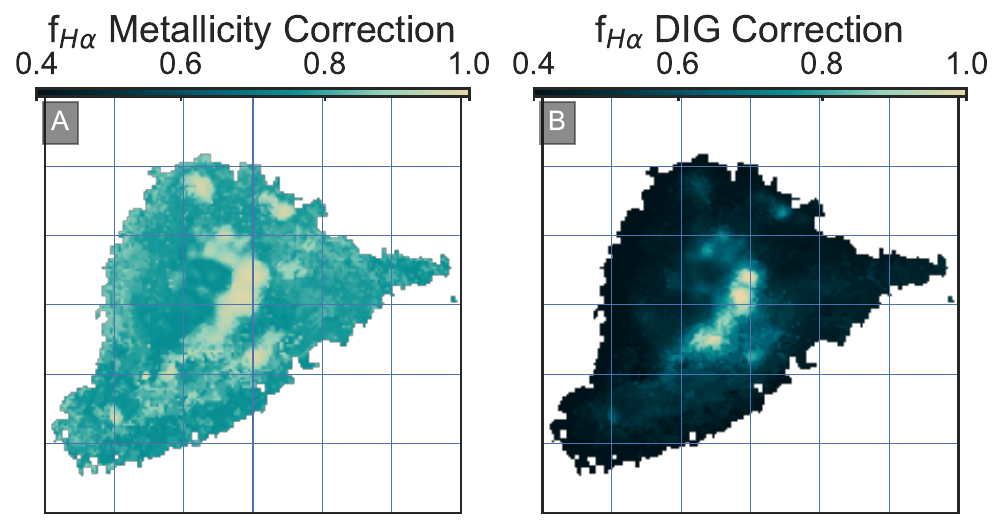}
 \caption{Correction maps applied to our \ha\ emission line maps. Panel A shows the correction which accounts for the \ha\ + [NII] blended flux using metallicity estimations. Panel B shows the correction accounting for \ha\ emission due to DIG. By multiplying the \ha\ map by these correcting factors we end up with a deblended \ha\ map composed of emission due to star formation.}
 \label{fig:corr}
\end{figure*}

\editone{The flux in each pixel of our \ha\ maps is actually \ha\ + [NII], as these lines will always be blended in the grism spectra. To account for this we estimate the [NII] to \ha\ ratio by deriving the spatially resolved metallicity. We do this by first deriving the ionization map using S32 ([SIII]$\lambda\lambda$ 9069, 9532/[SII]$\lambda\lambda$ 6717, 6731), and then deriving metallicity using S23 ([SIII]$\lambda\lambda$ 9069, 9532 + [SII]$\lambda\lambda$ 6717, 6731 / \ha) via equations from \cite{kewl19}. Note that [SIII]$\lambda$ 9069 falls outside of the NIRISS F200W coverage, so to convert the [SIII]$\lambda$ 9532 line flux to the total doublet line flux we use the [SIII]$\lambda$ 9532 / [SIII]$\lambda$ 9069 flux ratio of $\sim$ 2.5 \citep{oste06,taya19}. As [NII] / \ha\ is a metallicity indicator we can then use the S23 metallicity to estimate [NII] / \ha\ using equations from \cite{kewl19}. This gives us a way to estimate [NII] / \ha\ that accounts for metallicity gradients. Figure \ref{fig:corr} Panel A shows the spatially resolved fraction of \ha\ ($f_{H\alpha}$) using this metallicity correction. We plan to further analyze the spatially resolved metallicity map of the QMP in a future paper. }

\editone{Last, we need to account for \ha\ emission from sources other than star formation. The first alternate source we look at is diffuse ionized gas (DIG, \cite{mart97, sand17, zhan17}). DIG can account for a significant amount of \ha\ emission, accounting for 30$\%$ - 60$\%$ of \ha\ in local spiral galaxies \citep{zuri00, oey07} and is shown in \cite{sand17} to be correlated with the \ha\ surface density. Using equation 24 from \cite{sand17} we estimate the fraction of \ha\ originating from DIG (f$_{DIG}$) and derive from it our DIG correction seen in Figure \ref{fig:corr} Panel B (shown here as f$_{H\alpha}$ or 1 - f$_{DIG}$), where we see that in the outskirts of the galaxy as much as 60$\%$ of \ha\ is from DIG. DIG also affects metallicity estimations as metallicities are calibrated using HII region models, which do not match the conditions of DIG. Therefore we also apply a correction to our [SII] maps using equations from \cite{sand17}. We do not apply any corrections to our [SIII] map as DIG primarily enhances low-ionization emission lines \citep{sand20}. Therefore, our metallicity and DIG corrections are intertwined. To estimate these corrections we iterated over the derivation until the maps converged.}

\editone{Other sources of possible \ha\ emission are active galactic nuclei (AGN) and shocks. Strong AGN have a very distinctive look in grism emission line maps. When no AGN templates are included the continuum and emission are not properly modelled creating a cross pattern in the emission line map (it is cross as NIRISS uses orthogonal dispersion angles). We do not see this cross pattern in our maps and therefore do not believe there is a strong AGN present. There may be a weak AGN, but we do not have the necessary emission lines needed to create a BPT diagram to identify AGN emission \citep{bald81}, and so do not correct for this possibility. As for shocks, we currently do not have the data needed to estimate the proportion of \ha\ due to purely shocks as kinematic information and high-resolution spectra would be necessary to make this distinction \citep{kewl19}. For this proof of concept work, we assume no additional \ha\ emission from AGN or shocks. This assumption may lead to biases in our results, and we plan to add AGN and shock templates in our larger sample studies and will study the spatially resolved effects of AGN and shocks in greater detail in future work.
}

\subsection{H$\alpha$ emission line map \label{sec_ha_map}}
Figure \ref{fig:maps} Panel D shows the (dust, metallicity, DIG, and lensing corrected) H$\alpha$ emission line surface density map for the QMP, PSF-matched to the JWST/NIRCam F444W filter. The map is shown in log space to emphasize the H$\alpha$ emission in the face-on disk. We see that the red edge-on galaxy shows less H$\alpha$ flux (5.8  $\pm$ 0.1 $\times$10$^{-17}$ erg/s/cm$^2$) when compared to the blue face-on galaxy (10.0  $\pm$ 0.04 $\times$10$^{-17}$ erg/s/cm$^2$). We also see that a large amount of the H$\alpha$ flux is concentrated in the clumpy star-forming regions (4.1 $\pm$ 0.02 $\times$10$^{-17}$ erg/s/cm$^2$), accounting for 41$\%$ of the total \ha\ flux. One caveat we would like to add is that the red edge-on galaxy is highly obscured and we may be missing some H$\alpha$ flux. As we have no better dust measurement (H$\beta$ falls below our wavelength range and therefore we cannot correct using the Balmer decrement), we move forward assuming our dust maps do provide sufficient corrections. 

\subsection{SFR}
With the H$\alpha$ map in hand, we can begin to study additional properties of the galaxy. Using the relationship from \cite{kenn94} we calculate the SFR from the H$\alpha$ flux, which details recent star formation (timescale of approximately $\sim$ 10 Myr). The resulting SFR$_{H\alpha}$ map is shown in Figure \ref{fig:maps} Panel D. Here we see that the regions with the highest SFRs in the blue face-on galaxy are concentrated in the inner chain of clumpy regions. 

In total, the SFR of the blue face-on galaxy is 6.0 $\pm$ 0.2 M$_\odot$/yr (sSFR -9.3 $\pm$ 0.1 yr$^{-1}$) while the red edge-on galaxy has an SFR of 3.6 $\pm$ 0.3 M$_\odot$/yr (sSFR -9.2 $\pm$ 0.2 yr$^{-1}$). Breaking the galaxies into their components we see that the red edge-on galaxy, $\sim$ 87$\%$ of its star formation occurs in its disk and $\sim$13\% in its bulge. For the blue face-on galaxy, $\sim$ 54$\%$ of its star formation occurs in its disk, while $\sim$ 41$\%$ of its star formation happens in its clumpy star-forming regions, with the bulge accounting for the remaining $\sim$ 5$\%$. 

\section{Discussion}

\subsection{Star Formation burstiness from \ha-to-UV flux ratios
\label{sec:burst}}
\begin{figure*}
 \includegraphics[width=\linewidth]{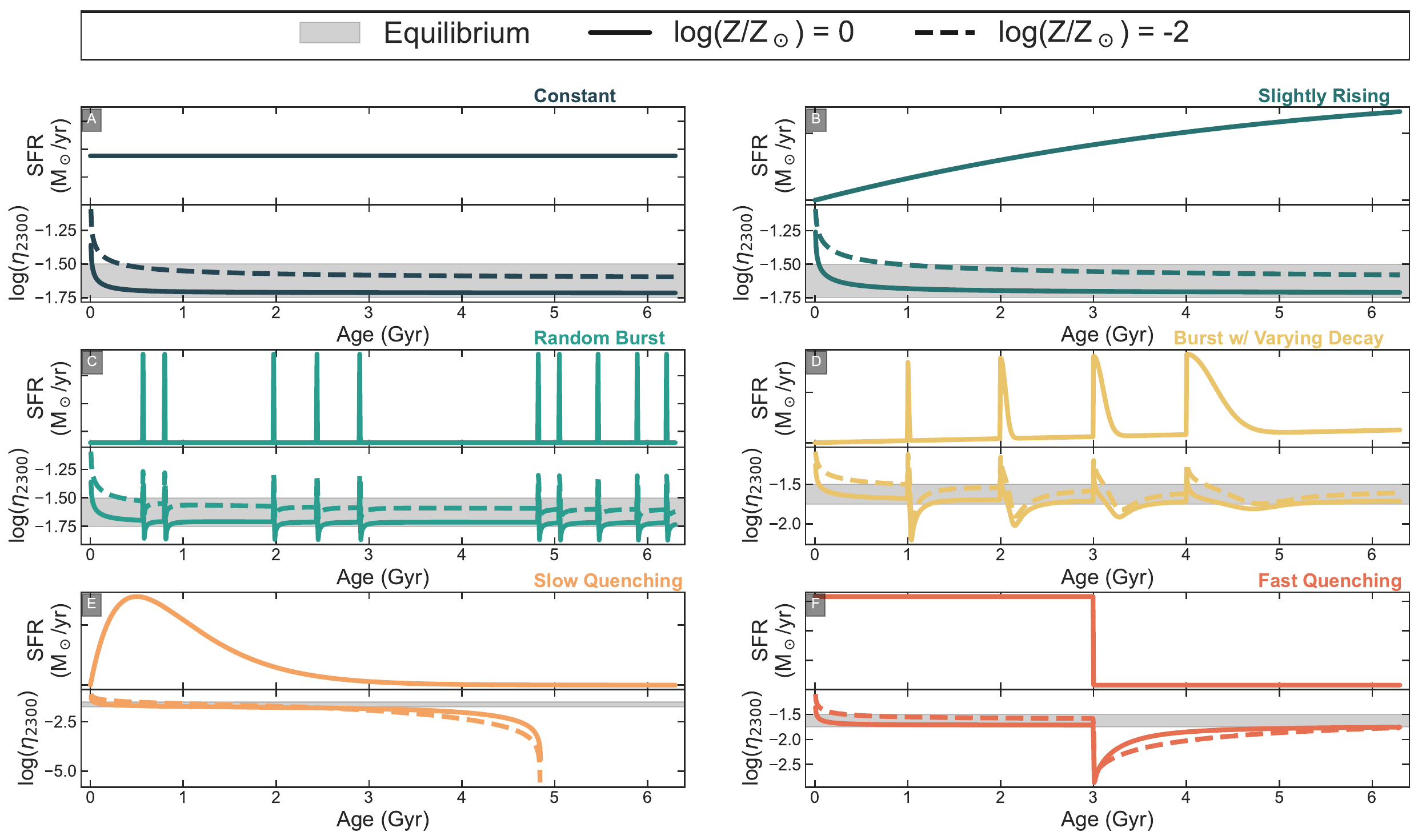}
 \caption{The connection between a galaxies SFH and \leta. In each of the panels, the top sub-panel shows the SFH we are examining and on the bottom sub-panel, we show the evolution of \leta\ for galaxies with log(Z/Z$_\odot$) = -2 (dashed line) and log(Z/Z$_\odot$) = 0 (solid line) along with the region of equilibrium (grey shaded region) derived using FSPS models. When $\eta$ is above the equilibrium region the SFH is in a bursting phase, in the equilibrium region the SFH is either flat or slightly rising or transitioning from bursting to quenching, and below the equilibrium region the SFH is in a quenching phase (either long-term or short-term quenching).  }
 \label{fig:UVHaV}
\end{figure*}

One of the main properties we are interested in studying for the QMP is the spatially resolved burstiness of star formation. We do this by taking the ratio of the dust-corrected H$\alpha$ and UV$_{2300}$ (HST/ACS F435W - restframe 2300 \AA) maps (Figure \ref{fig:maps} Panels D and G show the dust and lensing-corrected H$\alpha$ and UV$_{2300}$ maps respectively). \editone{The largest potential source of error in our burstiness calculations is our dust correction, given the wavelength separation and the fact that the UV is continuum while \ha\ is an emission line. Please refer to Section \ref{sec:corr} for our dust correction methods and any associated caveats.} Both H$\alpha$ and UV fluxes are commonly used to estimate SFR at different timescales \citep{kenn94, kenn12}. Stemming from gas around massive short-lived stars, H$\alpha$ is sensitive to timescales in the tens of millions of years. In contrast, the UV continuum flux originates from longer-lived stars and is sensitive to star formation on timescales up to hundreds of millions of years. When a burst of star formation occurs, we see that H$\alpha$ increases quickly, while the UV continuum flux is slower to respond. 

\begin{equation}\label{eq:eta}
    \eta_{2300} = F_{H\alpha} / \lambda F_{\lambda,UV}.
\end{equation}
Figure \ref{fig:UVHaV} shows how the $\eta_{2300}$
ratio responds to several illustrative star formation histories and suggests that we can use this flux ratio to constrain the time-variability of star formation \citep{lee09,weis12,emam19,meht23,asad23, asad24}. The subscript 2300 in Eq.~\ref{eq:eta} indicates that the measurement of the UV continuum is at rest-frame $\sim$ 2300\AA. This is because it is crucial to specify the wavelength at which the rest-frame UV flux is measured as the numerical values of $\eta$ depend upon it despite the overall qualitative behaviour remaining unchanged.

All panels in Figure \ref{fig:UVHaV} are structured so that the onset of star-formation is at Age $=$ 0 with the top plot of each panel showing the SFH and the bottom panel showing the evolution of \leta. \editone{Each of the \leta\ tracks  derived using FSPS with the settings stated in Section \ref{sec_BB}, assuming no dust and an ionization of log(U) = -3.} In all panels of Figure \ref{fig:UVHaV} we show a grey region defined as the \leta\ equilibrium (-1.75 $<$ \leta\ $<$ -1.55 ), which is roughly where the ratio resides during periods of constant or slightly rising star formation, as can be seen in Panels A and B respectively. The size of this region was determined by varying the metallicity of the models, with the lower bound determined by log(Z/Z$_\odot$) $=$ -2 (dashed line in bottom plots of each panel), and the higher bound determined by log(Z/Z$_\odot$) $=$ 0 (solid line in bottom plots of each panel). 

\editone{Panels A and B show that equilibrium can occur with a constant or slightly rising SFR, indicating that to leave the equilibrium region a large sudden change in SFR is required. Panels C and D show such cases as these are bursty SFHs, with Panel C showing bursts with rapid rises and declines, while Panel D show bursts with rapid rises and slow declines. Panels C and D show that the rapid change in SFR will cause \leta\ to leave the equilibrium region while Panel D also shows that during the fall of a burst \leta\ can spend an extended time in the equilibrium region. Panel D also shows that the value of \leta\ is correlated to how rapid the burst and subsequent fall-off are. Panels E and F show that being below the equilibrium region (the quenching phase) can occur without a burst and can happen through slow quenching (Panel E) and rapid quenching (Panel F). These SFHs indicate that being above the equilibrium region (bursty phase) only occurs when a burst happens while being below the equilibrium region can occur during the fall of a burst or long-term quenching. Therefore to understand the full context of \leta\ we must combine these results with spatially resolved SFHs.}


\begin{figure*}
 \includegraphics[width=18cm]{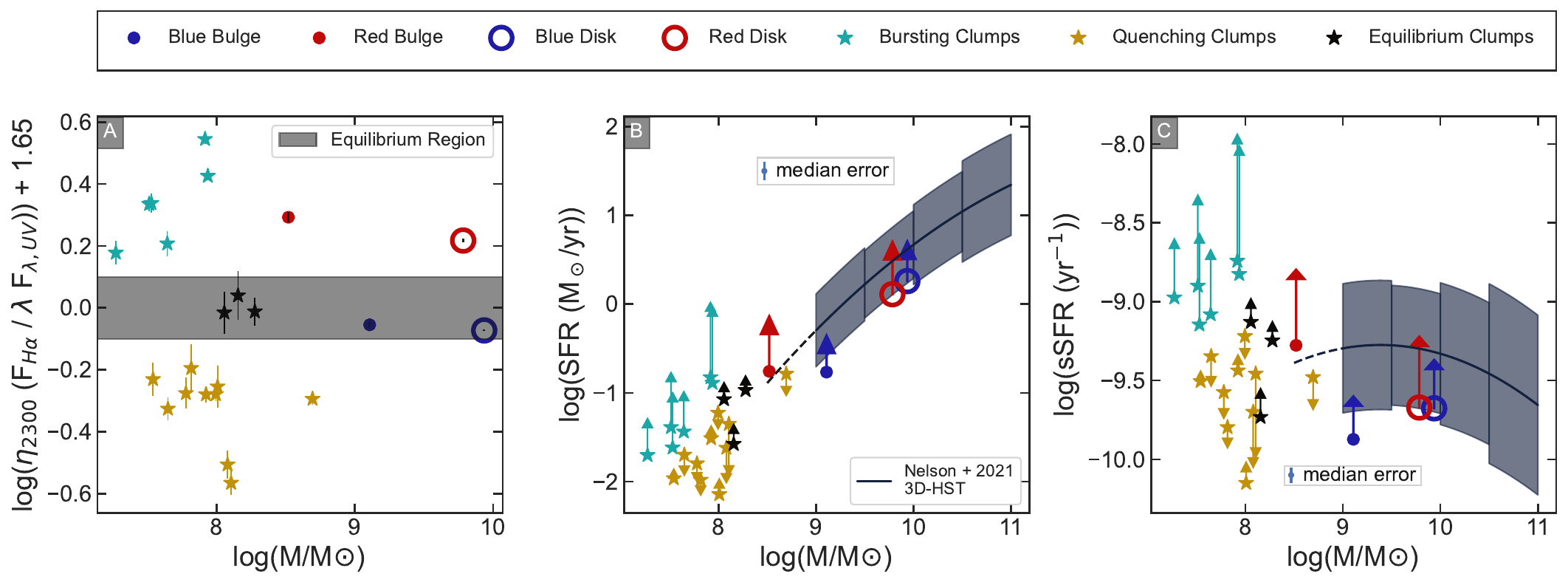}
 \caption{SFH properties vs stellar mass for each of the regions indicated in Figure \ref{fig:maps} Panel B (\editone{defined in Section \ref{sec:qmp}}, blue and red galaxy bulges and disks as well as the bursting and non-bursting clumps). Here the data points are plotted as arrows with the beginning of the arrow indicating the broadband measurement and the end of the arrow indicating the H$\alpha$ derived property. The span of the arrow indicates how large the change is over the relative timescales ($\sim$ 100 Myr) while the directions tell us if that value is rising or falling. Panel A shows the relationship between \leta\ and mass with the region of equilibrium shown in grey. Panel B shows the SFR-mass relationship with the star-forming main sequence from \citet{nels21}. In Panel C we show the sSFR-mass relationship including the star-forming main sequence from \citet{nels21}.  This figure shows that regions with bursting star formation do not necessarily need to be above the main sequence.}
 \label{fig:reg_m}
\end{figure*}
\subsection{Spatially Resolved Burstiness}

Figure \ref{fig:maps} Panel H shows the spatially resolved burstiness (\leta, see Section \ref{sec:burst}) of the QMP.  Here we have added 1.65 to \leta\ to set the equilibrium region to zero, with red indicating quenching, blue indicating bursting, and black indicating being in equilibrium. As noted in Section \ref{sec:burst} sitting above the equilibrium is associated with bursting star formation while values below the equilibrium correspond to quenching whether that be long-term quenching (i.e. exponentially decaying star formation) or short-term quenching (downturn of the SFH associated with the end of a burst). 

In Figure \ref{fig:maps} Panel H we see that the majority of the blue face-on galaxy is quenching (65$\%$ of the pixels below the equilibrium, 18$\%$ in equilibrium, and 17$\%$ sitting above - with 59$\%$ of the mass below the equilibrium, 19$\%$ in equilibrium, and 22$\%$ above). In contrast, the red edge-on galaxy is bursting with star formation (10$\%$ of the pixels below the equilibrium, 19$\%$ in equilibrium, and 71$\%$ sitting above - with 8$\%$ of the mass below the equilibrium, 24$\%$ in equilibrium, and 68$\%$ above), and the blue galaxy is therefore either experiencing more long-term quenching or post-burst quenching. \editone{\cite{asad24} showed that the \leta\ offset below the equilibrium is correlated with how rapid the quenching is, therefore the quenching clumps are experiencing a more rapid quenching event than the disk as the quenching clumps have an overall lower \leta\ measurements. It may be that clumps that are in equilibrium/quenching may have entered a post-burst phase and it may be that the reason we see smaller \leta\ values in the non-bursty regions is that these clumpy regions can more efficiently use up their fuel, while the disk is inefficient and therefore is slowly quenching. Larger samples would be needed to say if that was a common trait, something we will address in future work.} 
In Figure \ref{fig:maps} Panel I we show another parameterization of burstiness using the SFR$_{H\alpha}$ to the broadband SFR (SFR$_{BB}$) ratio (\sfrr). SFR$_{BB}$ was calculated for each region by taking the posterior SFH of that region, sampling from it 1000 times and then calculating the mean SFR over the final 100 Myr as this timescale is similar to the timescale of the UV emission. Panel I is shown using the same colour scheme as Panel H where red indicates quenching, blue indicates bursting, and black indicates being in equilibrium (i.e. SFR$_{H\alpha}$ and SFR$_{BB}$ having similar values). The results using \sfrr\ are the same as with \leta\ where we find that the blue face-on galaxy mostly sits in or below the equilibrium, and the red edge-on galaxy is mostly bursting. 

\begin{figure*}
 \includegraphics[width=18cm]{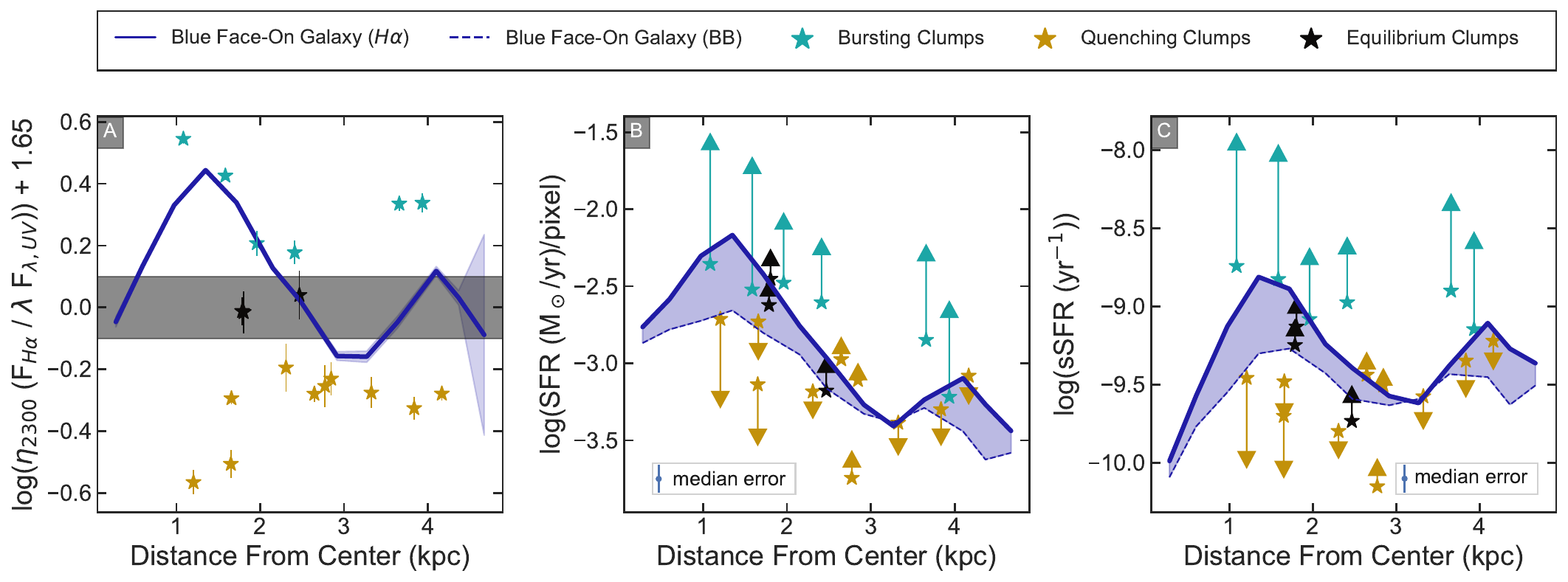}
 \caption{The change in SFH properties in reference to the distance from the centre of mass of the galaxy. Only the blue face-on galaxy is shown here as the red face-on galaxy is highly inclined. \editone{In all panels bursting/equilibrium/quenching clumps are shown as cyan/black/yellow arrows (with the direction of the arrow indicating the direction of change). The blue face-on galaxies disk is shown as a solid blue line for values derived from our \ha\ maps and a dashed line for values derived from the broadband data with the solid blue region showing the gap between the two values.} Panel A shows the evolution of \leta, Panel B shows the evolution in SFR and Panel C shows the evolution of sSFR. \editone{By comparing the panels we see that burstiness is best correlated with sSFR, while not being well correlated with SFR or distance from the centre.}}
 \label{fig:reg_d}
\end{figure*}

\subsection{Spatially Resolved Properties}
\subsubsection{Stellar Mass Dependence}
In Figure \ref{fig:reg_m} we show how the properties of \leta\, SFR, and sSFR correlate to mass for the regions shown in Figure \ref{fig:maps} Panel B \editone{(defined in Section \ref{sec:qmp})} with the blue and red galaxy bulges shown as dots (blue and red respectively), the blue and red disk shown as rings (blue and red respectively), and the bursting/equilibrium/quenching clumps shown as cyan/black/yellow stars. 

Panel A of Figure \ref{fig:reg_m} shows \leta\ vs stellar mass. We see no relationship between burstiness and stellar mass for the multiple regions. What we find is that the bursting clumps and the red disk and bulge sit above the equilibrium region, while the equilibrium clumps and blue disk/bulge sit in the equilibrium region, and only the quenching clumps sit below equilibrium. \editone{A question which arises is why the blue disk is in equilibrium when the majority of it is quenching (60$\%$ of its mass is below equilibrium). The reason for this is that the bursting regions within the blue disk exhibit significantly higher \ha\ emission, which elevates the \leta\ from a quenching state to equilibrium. This highlights a drawback of integrated \leta\ measurements, as a bursting region can easily outshine the rest of the galaxy. This effect is particularly noticeable in larger regions that contain a mixture of bursting and non-bursting areas. However, it is less problematic in smaller, more uniform regions, such as the star-forming clumps.} 

Figure \ref{fig:reg_m} Panel B shows the SFR - mass relationship for the regions along with the galaxy star-forming main sequence from \cite{nels21}. Each of the regions is marked with an arrow, with the arrow's starting point being the SFR measured from the broadband fits (correlating to a timescale of $\sim$ 100 Myr) while the endpoint is the SFR derived from the H$\alpha$ emission (correlating to a timescale of $\sim$ 10 Myr). Therefore the direction of the arrows shows us if the region's SFR has been increasing or decreasing over the last $\sim$ 100 Myr. In Panel B we see that the blue and red disks sit in the main sequence, while the blue bulge sits below. We see a correlation between $\Delta$ SFR and location on the main sequence. Note that we do not have a fit for the main sequence below log(M/M$_\odot$) = 9, though relatively the bursting clumps have larger $\Delta$ SFR values and sit above the equilibrium and quenching clumps which either have minor rises in SFR, or have a negative $\Delta$ SFR.  

In Panel C of Figure \ref{fig:reg_m} we show the sSFR - mass relationship along with the star-forming main sequence from \citet{nels21}. We find that the regions of the QMP mostly fall into the category of being star-forming with log(sSFRs) between -10.0 and -8.0 (yr$^{-1}$). \editone{One interesting takeaway from Panel B is that the bulge of the red galaxy has a higher sSFR than the disk. This may indicate that the red edge-on galaxy is experiencing a starburst induced by a possible interaction between the galaxies.}

\editone{When we combine the findings from the integrated values in all panels, it becomes clear that there is no correlation between a region’s position on the main sequence and \leta\ (therefore burstiness) as we see both quenching and bursting galaxies in the main sequence.}


\subsubsection{Dependence on Distance From Galaxy Centre}

In Figure \ref{fig:reg_d} we show the change in \leta, SFR, and sSFR with the distance from the galaxy centre (bulge). Here we focus on the blue face-on galaxy as we have not corrected the red edge-on galaxy for inclination. In each panel the bursting/equilibrium/quenching regions are shown as cyan/black/yellow arrows, the blue face-on galaxy disk (with the clumps omitted) is shown in blue with the solid line indicating values derived from H$\alpha$, and the dashed line for values from the broadband (corresponding to 10 and 100 Myrs respectively). The distances of each pixel have been corrected for lensing (see Appendix A).

In Figure \ref{fig:reg_d} Panel A we see how burstiness evolves with distance from the centre. In both the disk and clumps, we see larger offsets (both bursting and quenching) closer to the centre of the galaxy. To determine how burstiness correlates with distance from the centre we will need to utilize a larger sample of galaxies, something we plan to do in future work. 

In Panel B of Figure \ref{fig:reg_d} we see that the SFR for the blue galaxy rises outwardly in the central 1 kpc and then steadily drops going towards the outskirts of the galaxy. We also see that the clumpy star-forming regions in the outskirts have lower SFRs than those towards the centre of the galaxy. When comparing SFR$_{H\alpha}$ to SFR$_{BB}$ for the blue face-on galaxy we see that the regions with the largest increases are associated with bursting clumps with larger \leta\ values (as seen in Panel A). 

Figure \ref{fig:reg_d} Panel C shows how sSFR changes with distance from the centre, where we see that in the inner 2 kpc sSFR steadily rises from the core of the galaxy and peaks at $\sim$ 2 kpc (the approximate location of the inner chain of star-forming clumps), then lays mostly flat the rising in the outskirts ($\gsim$ 4.4 kpc). As for the star-forming clumps, we see that they are mostly flat in sSFR with distance.

\section{Conclusions}

We have presented the JWST/NIRISS case study of the spatial distribution of H$\alpha$ emission in a z=0.8718 interacting galaxy pair.  Our extraction of H$\alpha$ emission goes beyond the standard analysis that assumes a constant stellar population at all locations, and — exploiting the high-SNR data that JWST can deliver — allows for the variation of stellar populations throughout the galaxy, resulting in more accurate models of the NIRISS grism data. We find that:

\begin{itemize}
\item Multi-region models of BCGs better remove the contaminating spectra than their single-region counterparts. Using our approach it is now possible to analyze once unrecoverable contaminated spectra. 

\item Our multi-region models better model the JWST/NIRISS grism spectra, with residuals that are unbiased, unlike the single-region approach.

\item We find that the single-region emission line maps are heavily contaminated with poorly subtracted continuum and emission line flux. These maps find 50$\%$ more flux than our multi-region emission line maps, though this extra flux is all likely contamination.

\end{itemize}

This first illustration of our new technique is applied to a complex system of two potentially interacting galaxies. Using H$\alpha$/UV ratio ($\eta$) maps, which characterize the burstiness of star formation by exploiting the different timescales of H$\alpha$ (10 Myr) and UV (100 Myr) flux, we interpret our data as showing the following trends:  

\begin{itemize}
\item The higher-mass blue face-on galaxy is mostly quenching with a few patches or bursty star formation given their H$\alpha$/UV ratios. 

\item  Notably among the 20 star-forming clumps, 10 have low \leta\ ratios that suggest star formation in these clumps may have already peaked and has started to decline.

\item In contrast to the blue face-on galaxy, the lower-mass red edge-on galaxy has consistently higher \leta\ values, suggesting that it is currently bursting with star formation.  

\item When comparing \leta\ to our other burstiness indicator (\sfrr) we find that the two indicators are in agreement that quenching/equilibrium regions in the galaxy align spatially with clumpy star-forming regions. 

\item We speculate that the reason that the quenching clumps have lower overall \leta\ values is that these regions may more efficiently use up their fuel, and therefore have entered a post burst phase. This process may take longer in the disk. 

\end{itemize}

Additionally we compare spatially resolved SFRs from H$\alpha$ emission and broadband SED fits and find:
\begin{itemize}
\item The red edge-on galaxy's SFR has increased over the last 100 Myr, combined with our \leta\ results this indicates that the galaxy is possibly experiencing a starburst as its bulge has a higher sSFR than its disk.


\end{itemize}

We have demonstrated the power of spatially-resolved NIRISS spectroscopy to study the spatial distribution of star formation properties in distant galaxies using our new technique for extracting emission line maps in the era of high-SNR JWST slitless grism observations. Several works at low redshift (z $<$ 0.2) have studied when/where star formation happens using large samples of IFU data \citep{gonz16,belf18,medl18,avil23}. Our future work will extend the technique we introduced in this paper to larger, statistically significant, and complete samples of galaxies to determine when, where, and how star formation is happening at Cosmic Noon. 

\section*{Acknowledgements}

This research was enabled by grant 18JWST-GTO1 from the Canadian Space Agency and a Discovery Grant from the Natural Sciences and Engineering Research Council of Canada. MB acknowledges support from the ERC Grant FIRSTLIGHT, Slovenian national research agency ARRS through grants N1-0238 and P1-0188, and the program HST-GO-16667, provided through a grant from the STScI under NASA contract NAS5-26555. The authors acknowledge the Texas A$\&$M University Brazos HPC cluster and Texas A$\&$M High-Performance Research Computing Resources(HPRC, http://hprc.tamu.edu) that contributed to the research reported here.
This research used the Canadian Advanced Network For Astronomy Research (CANFAR) operated in partnership with the Canadian Astronomy Data Centre and The Digital Research Alliance of Canada with support from the National Research Council of Canada the Canadian Space Agency, CANARIE and the Canadian Foundation for Innovation.
%

\section*{Data Availability}


Raw JWST data used in this work will be available from the {\it Mikulski Archive for
Space Telescopes} (\url{https://archive.stsci.edu}), 
at the end of the 1-year proprietary time (doi: 10.17909/ph4n-6n76). Processed data products will be available on a similar timescale at \url{http://canucs-jwst.com}.



\bibliographystyle{mnras}
\bibliography{library} 




\appendix
\section{Lens Models \label{sec_lens}}

A strong lensing model is built using \texttt{Lenstool} \citep{knei93, jull07} and will be presented in a separate paper (G.~Desprez et al., in prep.). This model is leveraging the multiple images constraints from \cite{mahl19} and \cite{jauz19}, as well as new multiple images systems and redshifts obtained from the CANUCS-\textit{JWST} data. The model includes cluster-size mass halos and galaxy sizes, described as double Pseudo-Isothermal Elliptical (dPIE) profiles \citep{elia07}. Additionally, six clumps identified in the different images of the QMP are used as model constraints and allow us to produce a highly accurate lens model in the area of the QMP. The quality of our model can be ascertained from the fact that the average distance between the observed and predicted positions of the different multiple QMP clump images is 0.27\arcsec, whereas it is 0.44\arcsec\ for all the multiple images in the model. This indicates that the model is well-constrained, and particularly so around the QMP.



\bsp	
\label{lastpage}
\end{document}